\documentstyle[prb,twocolumn,aps]{revtex}

\begin{document}

\draft

\title{Comment on ``Inverse exciton series in the
optical decay of an excitonic molecule''}

\author{I. S. Gorban, M. M. Bilyi, I. M. Dmitruk,
and O. A. Yeshchenko}
\address{Department of Experimental Physics,
Physics Faculty, Kyiv Taras Shevchenko University, \\ 
6 Akademik Glushkov prosp., 252022 Kyiv, Ukraine \\
E-mail: yeshchen@expphys.phys.univ.kiev.ua}

\date{\today}
\maketitle

\begin{abstract}
Tokunaga {\em et al.} [Phys. Rev. B {\bf 59}, R7837
(1999)] claim the first successful observation of the  
inverse exciton $M$ series in the emission spectrum
of excitonic molecules. We assert that actually such a
series was observed as early as 1989 in the $\beta$-$ZnP_{2}$ 
crystal. We show that the objections of Tokunaga {\em et al.} against 
the biexciton nature of the inverse exciton series in 
$\beta$-$ZnP_{2}$ are ungrounded. In particular, their estimations 
for the ratio of the intensities of $M2$ and $M1$ emission lines in 
this crystal give two order smaller value because they do not take 
into account the reabsorption of the $M1$ line photons. We report the 
observation of additional inverse exciton $M^{\prime}$ series in the 
emission spectrum of excitonic molecules in $\beta-ZnP_{2}$.  
\end{abstract}


In the recent publication \cite{CritPap} Tokunaga {\em et
al.} have reported an observation of the inverse exciton
$M$ series in the emission spectrum of excitonic molecules
in $CuCl$ crystal. They claim that their experiment is the
first successful observation of such a series in semiconductors. 
The aim of our Comment is to show that the above claim is not actual. 

To our knowledge the first observation of the emission
$M$ series caused by two-electron radiative transitions in a
biexciton was reported by some authors of the present
Comment ten years ago. \cite{JETPLett} In Ref.\ \onlinecite{JETPLett} 
such a series was observed in $\beta -ZnP_{2}$ crystal (fig.\
\ref{Fig1}). At two-electron transition one exciton of the molecule
annihilates and the other stays in the ground state (line $M1$) or
passes to excited (lines $M2$, $M3$, ...) and ionized (line
$M\infty$) states. Later, we described inverse exciton series in 
$\beta-ZnP_{2}$ emission spectrum in Ref.\ \onlinecite{PSSEM}. 
However, the authors of Ref.\ \onlinecite{CritPap} disregarded above 
publications. Tokunaga {\em et al.} cited only the paper,
\cite{SSCTr} where the theoretical analysis of two-electron and
two-photon radiative transitions in an excitonic molecule was made. 
However, they have objected the biexciton nature of the inverse
exciton series in $\beta -ZnP_{2}$.

First their argument is that the ratio of the intensities of $M2$
and $M1$ lines of the series in $\beta -ZnP_{2}$ is too large for the
biexciton $M$ series: $I_{2}/I_{1} \sim 10^{-1}$. We would like to 
point an attention of the authors of Ref.\ \onlinecite{CritPap} to 
the following. Indeed, an experimental value of $I_{2}/I_{1}$ is 
$1.1\times 10^{-1}$.  Proceeding from the assumption of biexciton 
nature of the inverse exciton series in $\beta -ZnP_{2}$, we obtained 
theoretically in Ref.\ \onlinecite{SSCTr}: $I_{2}/I_{1} =
1.1\times 10^{-2}$. Numerical estimations of the authors of Ref.\
\onlinecite{CritPap} give $\sim 10^{-3}$ for $I_{2}/I_{1}$ in this
material. However, as we have pointed out in Ref.\ 
\onlinecite{SSCTr}, this disagreement between theory and experiment 
is clear enough.  At high excitation of the sample by powerful pulse
$N_{2}$ laser (excitation intensities $\sim 10^{6} W/cm^{2}$) the
concentration of $1S$ excitons is high enough. In addition to this,
in $\beta -ZnP_{2}$ the lowest exciton state is the forbidden state
of orthoexciton \cite{ExcConv} that could raise the concentration of
excitons in $1S$ state. Transitions from the $1S$ state of exciton 
into the molecule ground state cause reabsorption of $M1$ line 
photons ($he_{1S} + \hbar\omega _{M1} \rightarrow h_{2}e_{2}$) and 
corresponding decrease of this line intensity. Concentration of
excitons in the $2S$, $3S$, ... states does not increase considerably 
because of their fast relaxation to the $1S$ state, and there is no
saturation of the lines $M2$, $M3$ and $M\infty$. This our argument
is based on the results of experimental study of the dependences of
the intensities of inverse series lines on the excitation intensity
$I_{exc}$ (fig.\ \ref{Fig2}). \cite{JETPLett} The second power
increase of the $M1$ line intensity with $I_{exc}$ slows done at high 
excitation levels, and this effect becomes stronger with increase of
the excitation intensity. Such saturation effect is not observed for
other lines of the series. Let us extrapolate the $M1$ line intensity 
dependence on excitation intensity to the values of $I_{exc}$ at 
which an experimental value of $I_{2}/I_{1}$ is $1.1\times 10^{-1}$ 
(fig.\ \ref{Fig2}).  Consequently, we obtain the value of 
$I_{2}/I_{1}$ in the case of the absence of reabsorption of $M1$ line 
photons and the respective absence of this line saturation. An 
extrapolation gives $I_{2}/I_{2} = 4.3\times 10^{-3}$. This value 
agrees with Ref.\ \onlinecite{CritPap} estimations ($\sim 10^{-3}$) 
which do not consider the reabsorption effect. Some disagreement 
between the value obtained from the extrapolation and our estimations 
made in Ref.\ \onlinecite{SSCTr} ($1.1\times 10^{-2}$) is possibly 
due to rough wave function of biexciton that we have used in Ref.\ 
\onlinecite{SSCTr}. For other lines of the inverse series in $\beta
-ZnP_{2}$ good agreement between experimental and theoretical values
of the ratios of intensities takes place:  $(I_{3}/I_{2})_{exp} =
1.4\times 10^{-1}$ and $(I_{3}/I_{2})_{th} = 1.3\times 10^{-1}$;
$(I_{\infty}/I_{3})_{exp} = 4.3\times 10^{-1}$ and 
$((I_{4}+I_{5})/I_{3})_{th} = 3.9\times 10^{-1}$, where $I_{\infty} = 
\sum_{n=4}^{\infty} I_{n}$ is the total intensity of $M4$, $M5$, ...
lines which merge into the total $M\infty$line. In $CuCl$ the lowest 
exciton state is allowed. Biexcitons in Ref.\ \onlinecite{CritPap} 
were resonantly created by two-photon absorption method and, 
therefore, excitons would be created only at the optical decay of 
biexcitons. Due to these two facts, the concentration of excitons 
would be insufficient for reabsorption of $M1$ line photons, and this 
line would not be saturated, i.e. an agreement between experiment and 
theory would take place for all lines of the inverse series in 
$CuCl$. Such an agreement was reported in Ref.\ \onlinecite{CritPap}. 

$M$ series in $\beta -ZnP_{2}$ is observed at considerably higher
temperatures too (fig.\ \ref{Fig3}). This fact confirms high binding 
energy of the biexciton in $\beta -ZnP_{2}$ and rejects any possible 
impurity interpretation.  $\beta-ZnP_{2}$ is very remarkable in the 
following. If the direction of the wavevector of emitted photon ${\bf 
k}$ is parallel to axis $b$ of crystal (${\bf k} \parallel b$) and 
photon polarization is ${\bf E} \parallel a$, another inverse exciton 
$M^{\prime}$ series is observed in emission spectrum of this crystal 
(fig.\ \ref{Fig4}). This additional series is symmetric to $A$ series 
of the free $S$ orthoexciton which is observed in absorption spectrum 
at ${\bf E} \parallel a$ and ${\bf k} \parallel b$. \cite{Aser} Thus, 
$M^{\prime}$ series is due to the radiative transitions from the 
excitonic molecule ground state to the $S$ states of orthoexciton. 
Main $M$ series is symmetric to $C$ series of the free $S$ 
paraexciton which is observed in absorption and emission spectrum at 
${\bf E} \parallel c$. \cite{Aser} Respectively, $M$ series is due to 
the radiative transitions from the biexciton ground state to the $S$ 
states of paraexciton.

Authors of Ref.\ \onlinecite{CritPap} see the cause that the previous
attempts of biexciton $M$ series observation ``were unsuccessful'' in 
fact that ``the $M_{n \geq 2}$ lines are extremely weak in intensity, 
requiring for their observation a highly sensitive detection 
technique and a high-quality sample free from impurity emissions''.  
However, the fact of extremely weak intensities of $M$ series lines 
in $CuCl$ does not prove the impossibility of that in other materials 
these intensities can be considerably higher.  Indeed, in $CuCl$ 
intensities of $M_{n \geq 2}$ lines are extremely small, since in 
this material: $I_{2}/I_{1} = 1.4\times 10^{-4}$; $I_{3}/I_{1} = 
4.0\times 10^{-5}$; $I_{4}/I_{1} = 1.5\times 10^{-5}$. In $\beta 
-ZnP_{2}$ we have more than one order higher values: $I_{2}/I_{1} = 
4.3\times 10^{-3}$; $I_{3}/I_{1} = 6.0\times 10^{-4}$; $I_{4}/I_{1} = 
2.6\times 10^{-4}$. This fact simplifies the observation of $M$ 
series in $\beta -ZnP_{2}$. And finally, authors of Ref.\ 
\onlinecite{CritPap} asserted that our experimental data were not 
confirmed in a similar experiment by K. Kondo (K. Kondo, M.S. 
thesis, Okayama University, 1998). However, they have answered by 
themselves on this point having written that the high-quality samples
are required for biexciton $M$ series observation. Most likely, K.
Kondo just did not have the samples of the required quality.

The authors of Ref.\ \onlinecite{CritPap} determined the components
$C_{n}$ of the exciton excited states $n=2,3$, and $4$ in the
biexciton wave function in $CuCl$ from the relative intensities of
$M1$, $M2$, $M3$, and $M4$ lines. This allowed them to reconstruct
the internal molecule wave function. However, they did not say that
this idea was proposed in Ref.\ \onlinecite{SSCTr}. Only the
reabsorption of $M1$ line photons and the respective saturation of
this line did not allow us to reconstruct the biexciton wave
function in $\beta -ZnP_{2}$. However, taking for $I_{2}/I_{1}$ the 
extrapolation value obtained above, we can estimate the components 
$C_{n}$ of the exciton excited states in the biexciton wave function 
in $\beta -ZnP_{2}$ as following: $|C_{2}/C_{1}|^{2} = I_{2}/I_{1} = 
4.3\times 10^{-3}$; $|C_{3}/C_{1}|^{2} = I_{3}/I_{1} = 6.0\times 
10^{-4}$; $|C_{4}/C_{1}|^{2} = I_{4}/I_{1} = 2.6\times 10^{-4}$.

In conclusion, we assert that the inverse exciton $M$ series in the 
emission spectrum of excitonic molecules was observed for the first 
time in $\beta -ZnP_{2}$ crystal in 1989. \cite{JETPLett} Besides 
this series, we have observed in this crystal an additional inverse 
exciton $M^{\prime}$ emission series which is due to the radiative 
transitions from the biexciton ground state to the $S$ states
of orthoexciton.



\begin{figure}
\caption{(a) Biexciton $M$ series in emission spectrum of the
$\beta -ZnP_{2}$ at excitation intensity $\sim 10^{6} W/cm^{2}$
(pulse $N_{2}$ laser) and temperature $2 K$. (b) $S$ paraexciton
$C$ series in emission spectrum of the $\beta -ZnP_{2}$. Photon 
polarization: ${\bf E} \parallel c$. The inset presents the scheme of 
radiative transitions in biexciton causing the $M$ series.} 
\label{Fig1}
\end{figure}

\begin{figure}
\caption{Dependences of $M1$ and $M2$ line intensities on excitation 
intensity.}
\label{Fig2}
\end{figure}

\begin{figure}
\caption{Emission $M$ series in $\beta -ZnP_{2}$ at excitation 
intensity $\sim 10^{6} W/cm^{2}$ and temperature $77 K$. Photon 
polarization: ${\bf E} \parallel c$.} 
\label{Fig3} 
\end{figure}

\begin{figure}
\caption{Main biexciton $M$ series and additional biexciton
$M^{\prime}$ series in emission spectrum of the $\beta -ZnP_{2}$ 
at $I_{exc} \sim 10^{6} W/cm^{2}$ and $T = 2 K$.
${\bf k} \parallel b$, ${\bf E} \parallel a$.}
\label{Fig4}
\end{figure}

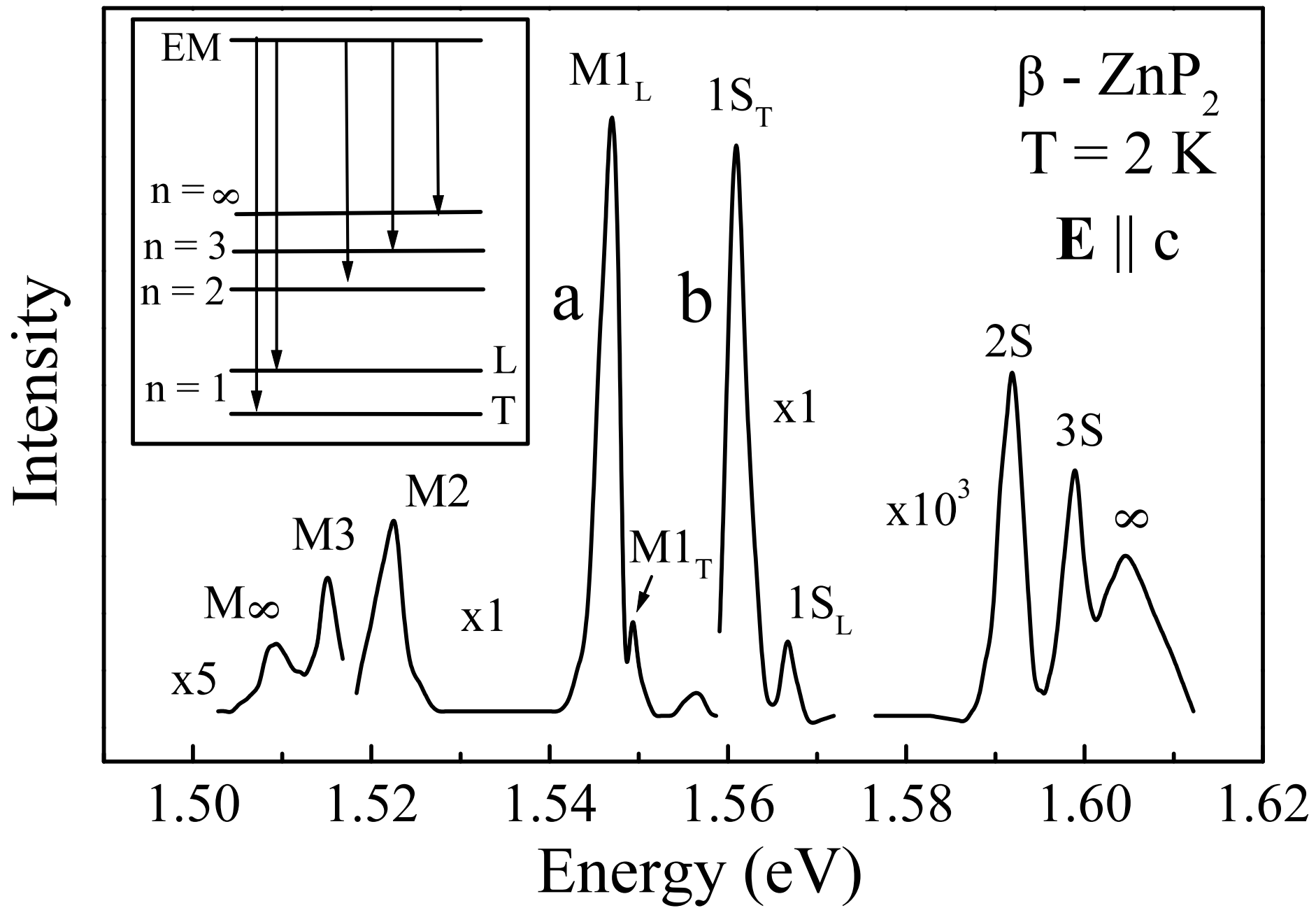
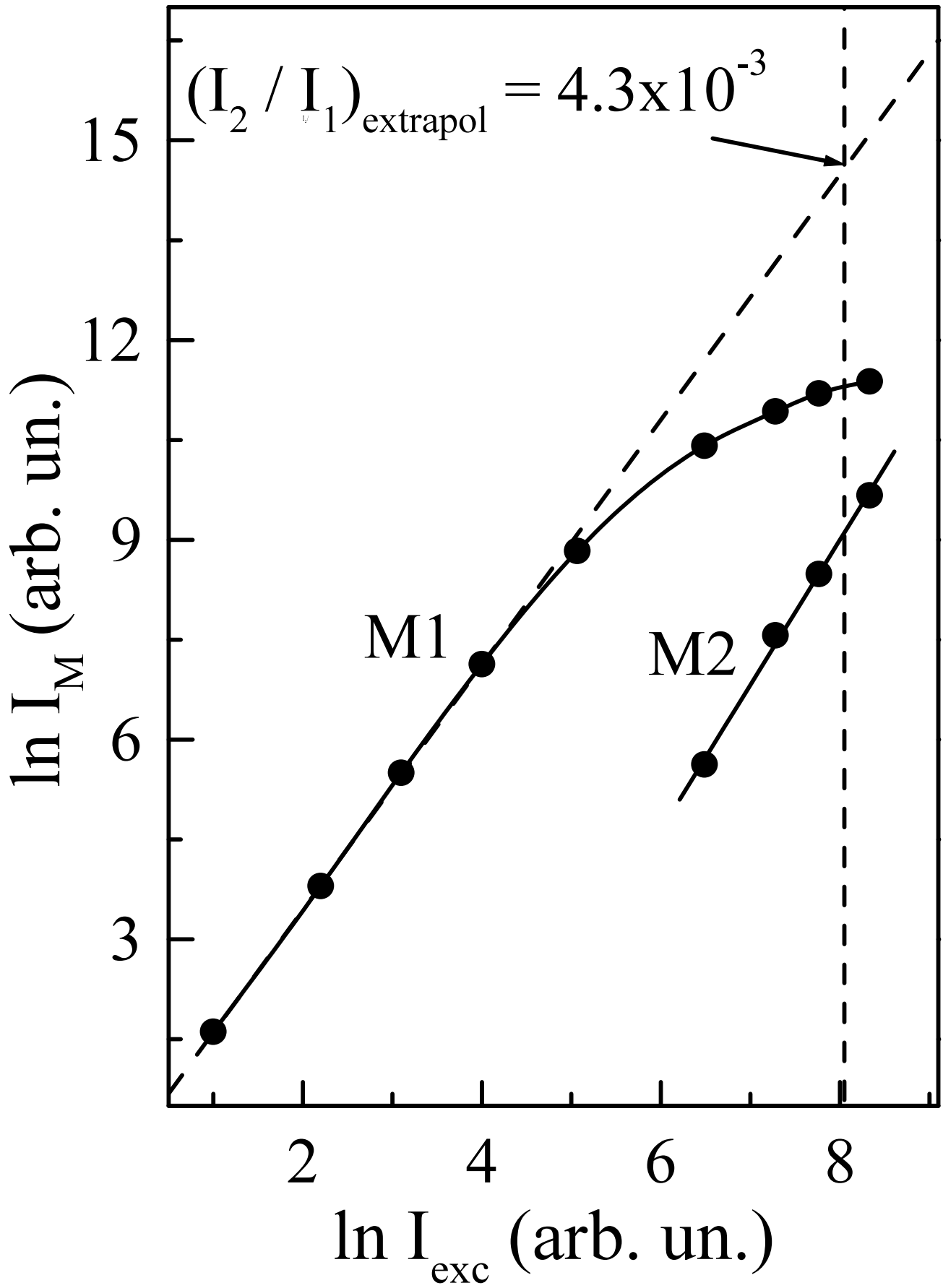
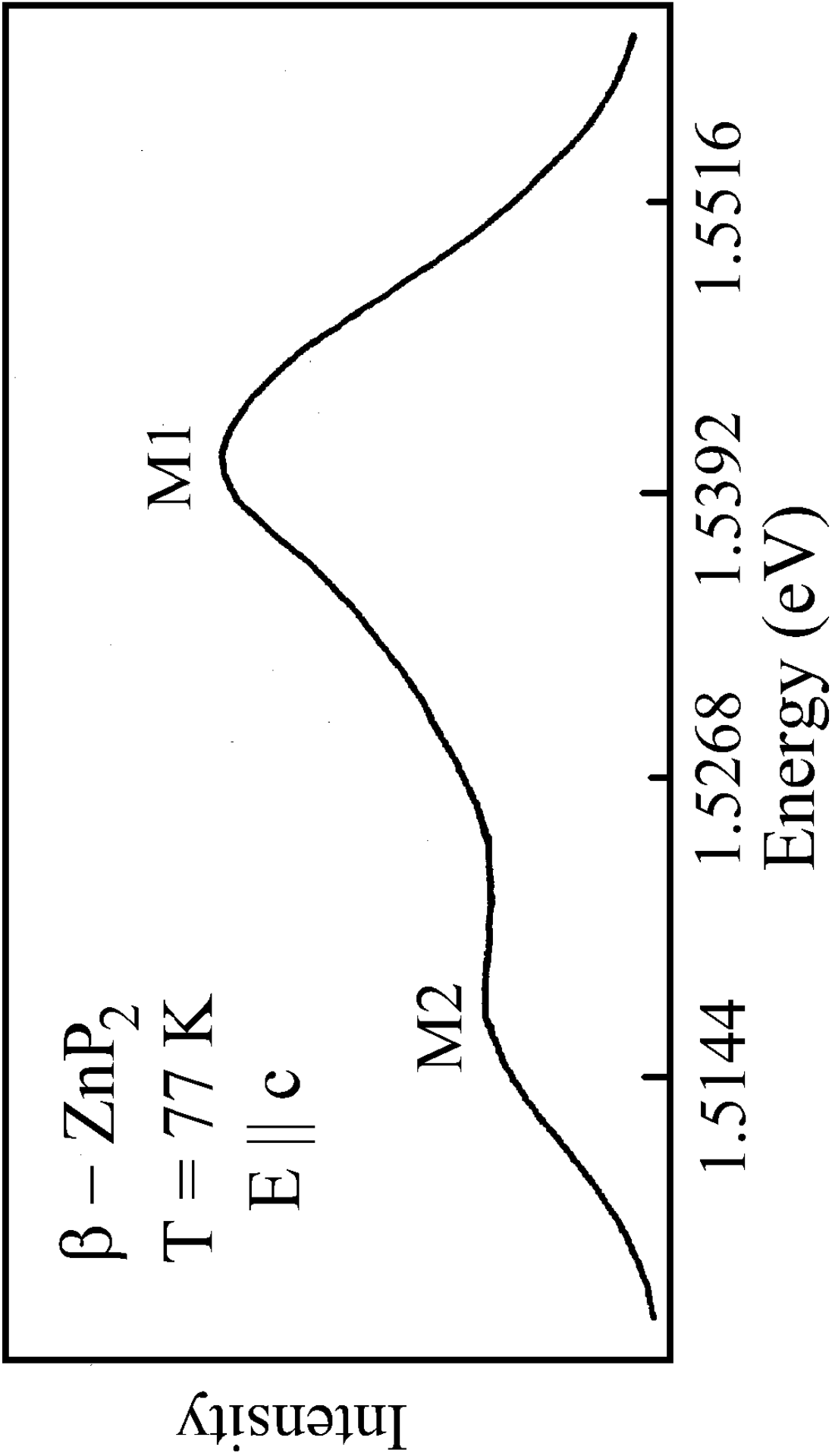
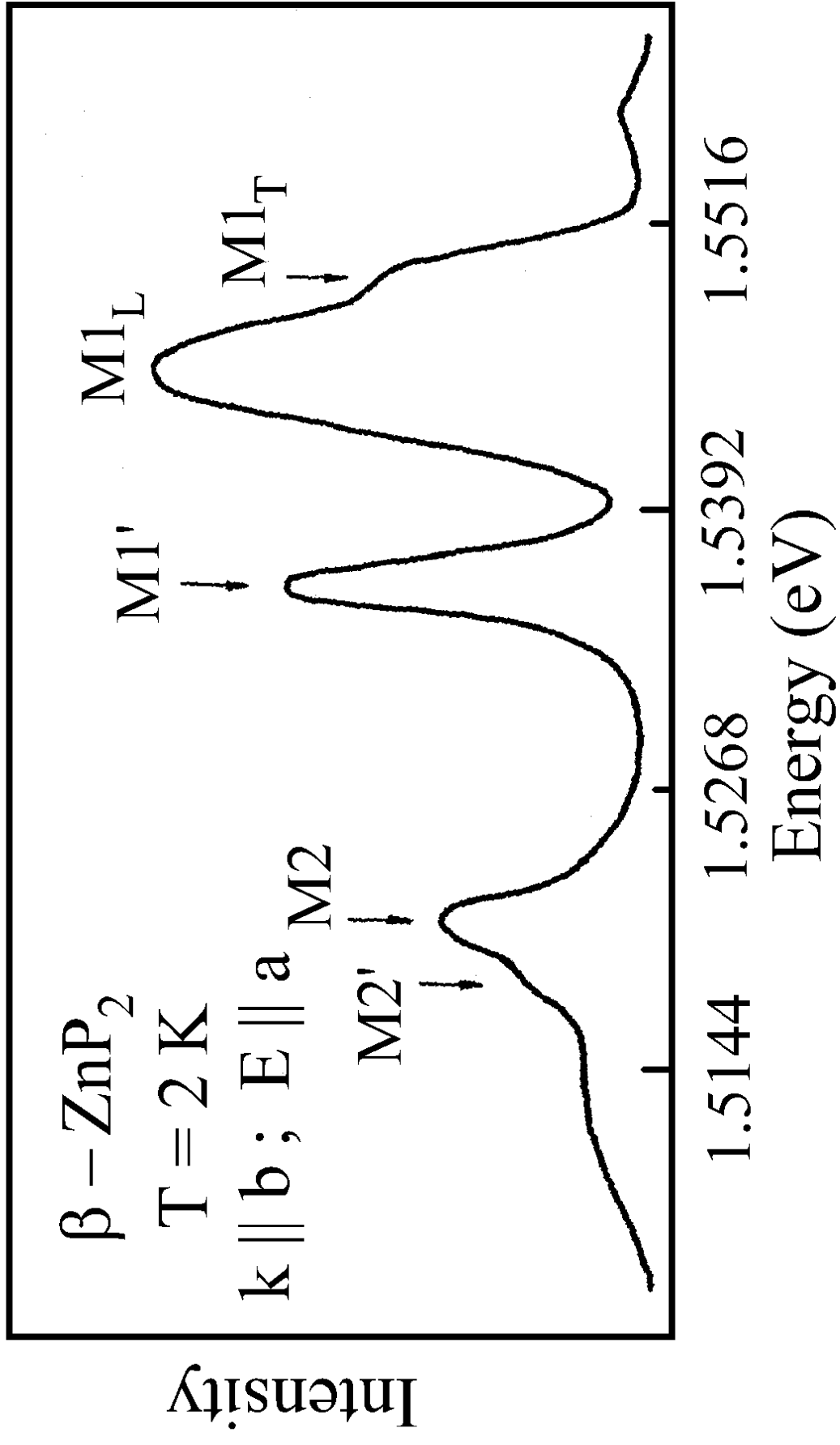
\end{document}